\documentclass[twocolumn]{aa}

\usepackage{graphicx}
\usepackage{txfonts}
\usepackage{color}

\begin{document}

\title{Can electron distribution functions be derived through the Sunyaev-Zel'dovich effect?}

\author{Prokhorov, D. A.\inst{1,2,3}, Colafrancesco, S.\inst{4,5}, Akahori, T.\inst{6}, Yoshikawa, K.\inst{7}, Nagataki, S.\inst{3},
 Seon, K.-I.\inst{1}}

\offprints{D.A. Prokhorov \email{phdmitry@stanford.edu}}

\institute{Korea Astronomy and Space Science Institute, Hwaam-dong,
Yuseong-gu, Daejeon, 305-348, Republic of Korea
             \and
Hansen Experimental Physics Laboratory, Kavli Institute for Particle
Astrophysics and Cosmology, Stanford University, Stanford, CA 94305,
USA
            \and
YITP, Kyoto University, Kyoto 606-8502, Japan
            \and
INAF - Osservatorio Astronomico di Roma
              via Frascati 33, I-00040 Monteporzio, Italy.
              Email: sergio.colafrancesco@oa-roma.inaf.it
             \and
              ASI
              V.le Liegi 26, Roma, Italy
              Email: Sergio.Colafrancesco@asi.it
            \and
Research Institute of Basic Science, Chungnam National University,
Daejeon, Republic of Korea
            \and
Center for Computational Sciences, University of Tsukuba, 1-1-1,
Tennodai, Ibaraki 305-8577, Japan
            }

\date{Accepted . Received ; Draft printed: \today}

\authorrunning{D.A. Prokhorov et al.}

\titlerunning{Can electron distribution functions be derived via the SZ effect?}

\abstract
{}
{Measurements of the Sunyaev-Zel'dovich (hereafter SZ) effect
distortion of the cosmic microwave background provide methods to
derive the gas pressure and temperature of galaxy clusters. Here
we study the ability of SZ effect observations to derive the
electron distribution function (DF) in massive galaxy clusters.}
{Our calculations of the SZ effect include relativistic
corrections considered within the framework of the Wright
formalism and use a decomposition technique of electron DFs into
Fourier series. Using multi-frequency measurements of the SZ
effect, we find the solution of a linear system of equations that
is used to derive the Fourier coefficients; we further analyze
different frequency samples to decrease uncertainties in Fourier
coefficient estimations.}
{We propose a method to derive DFs of electrons using SZ
multi-frequency observations of massive galaxy clusters. We found
that the best frequency sample to derive an electron DF includes
high frequencies $\nu$=375, 600, 700, 857 GHz. We show that it is
possible to distinguish a Juttner DF from a Maxwell-Bolzman DF as
well as from a Juttner DF with the second electron population by
means of SZ observations for the best frequency sample if the
precision of SZ intensity measurements is less than 0.1\%. We
demonstrate by means of 3D hydrodynamic numerical simulations of a
hot merging galaxy cluster that the morphologies of SZ intensity
maps are different for frequencies $\nu$=375, 600, 700, 857 GHz. We
stress that measurements of SZ intensities at these frequencies are
a promising tool for studying electron distribution functions in
galaxy clusters. }
{}

\keywords{galaxies: cluster: intracluster medium; relativistic
processes; cosmology: cosmic microwave background}

\maketitle

\maketitle

\section{Introduction}

Massive clusters of galaxies are the largest virialized objects in
the Universe bound by gravitation in the presence of dark matter.
Apart from galaxies and dark matter, galaxy clusters contain a hot
highly ionized plasma with temperatures up to 15 keV that emits in
X-rays: a continuum through bremsstrahlung and lines through
spontaneous decays of excitation states of ions. The hot
intracluster plasma can be fully described by the particle
distribution function (DF).

Equilibrium DFs are different in non-relativistic and relativistic
statistics. A thermal equilibrium DF is described by
Maxwell-Boltzman and Juttner functions in the framework of the
non-relativistic and relativistic theories, respectively (Landau \&
Lifshitz 1976). High-energy phenomena allow us to study thermal
mildly relativistic particle populations, such as mildly
relativistic electron populations in clusters of galaxies.

Inverse Compton (IC) scattering of hot free electrons in clusters of
galaxies on the cosmic microwave background (CMB) radiation field is
another effect that provides us with a method to study hot plasmas
in galaxy clusters, because IC scattering causes a change in the
intensity of the CMB radiation toward clusters of galaxies (the
Sunyaev-Zel'dovich effect, hereafter the SZ effect; for a review,
see Sunyaev \& Zel'dovich 1980), which depends on the details of the
electron distribution function in the cluster atmosphere (see, e.g.,
Colafrancesco et al. 2003).

A relativistically correct formalism for the SZ effect based on the
probability distribution of the photon frequency shift after
scattering was given by Wright (1979) to describe the Comptonization
process of soft photons by mildly relativistic plasma. Relativistic
corrections for the SZ effect allow us to measure the temperature of
intracluster plasma (see, e.g., Pointecouteau et al. 1998; Hansen et
al. 2002) and have been studied both analytically (Colafrancesco et
al. 2003, Colafrancesco \& Marchegiani 2010) and by means of
numerical simulations (Prokhorov et al. 2010a).

There are different methods proposed so far to derive a plasma
temperature by means of the SZ effect, which are based on the
measurement of one of following quantities: the shift of the
crossover frequency (see Rephaeli 1995), the intensity slope around
the crossover frequency (see Colafrancesco et al. 2009), the wide
frequency spectroscopy of the SZ effect spectrum especially at high
frequencies (Colafrancesco \& Marchegiani 2010), and the ratio of
the SZ intensities at two frequencies (Prokhorov et al. 2010a). As
was noticed by Rephaeli (1995) the correct relativistic equilibrium
distribution is essential for the proper interpretation of
measurements of the SZ effect.

One of the methods to study the relativistic equilibrium DF of
particles is to perform fully relativistic molecular dynamics
simulations (see Cubero et al. 2007 for the 1D case; Montakhab et
al. 2009 for the 2D case; and Peano et al. 2009 for the 3D case).
However, to justify the use of the relativistic equilibrium DF in
astrophysics, an observational confirmation is required.

In this paper we propose a method to derive the relativistic
equilibrium DF that is based on multi-frequency measurements of the
SZ effect in massive clusters. This provides us with a more complete
analysis of the electron distribution in the velocity space than
that given by knowing only the temperature value,  and a method to
verify the relativistic equilibrium DF. We also check the validity
of using the Juttner DF as an appropriate approximation to the
universal (equilibrium)  electron distribution in massive merging
clusters.

Hard X-ray emission tails reported in Beppo-SAX and RXTE X-ray
spectra of some galaxy clusters (see Fusco-Femiano et al. 1999;
Fusco-Femiano et al. 2004; Rossetti \& Molendi 2004, for the Coma
cluster; Petrosian et al. 2006 for the Bullet cluster) were
interpreted as bremsstrahlung emission from non-thermal
subrelativistic electrons (e.g. Sarazin \& Kempner 2000) or from
thermal electrons with a Maxwellian spectrum distorted by a particle
acceleration mechanism (Blasi 2000; Liang et al. 2002), or from
thermal electrons with a Maxwellian spectrum with a high temperature
(e.g., Petrosian et al. 2006). Million \& Allen (2009) reported the
discovery of spatially extended, non-thermal-like emission
components in Chandra X-ray spectra for five massive galaxy
clusters. Using Swift/BAT data, Ajello et al. (2010) have confirmed
the presence of  a hard X-ray excess from the  Bullet cluster. Note
that a possible contribution from inverse Compton emission of highly
relativistic electrons with Lorentz factor $\gamma\sim10^4$ to a
hard X-ray excess (for a review, see Rephaeli et al. 2008)
constrains our possibilities to study electrons DFs by using X-ray
observations. It has been suggested that a good test to check the
bremsstrahlung interpretation of hard X-ray tails is to use
multi-frequency measurements of the SZ effect (see Dogiel et al.
2007).

Spatially-resolved observations of the SZ effect can also provide
relevant information on the distribution of the electron plasma in
galaxy clusters with various temperatures. Prokhorov et al. (2010a)
found that the morphologies of the SZ intensity maps of a cool
galaxy cluster at frequencies of 128 GHz and 369 GHz are similar.
Here we perform numerical simulations of a hot merging galaxy
cluster to study the morphologies of the SZ intensity maps for a hot
merging galaxy cluster to explore the different morphologies for low
and high frequencies that depend on the importance of the SZ effect
relativistic corrections.

The layout of the paper is as follows. The dependence of the CMB
distortion caused by the SZ effect on the electron DF is considered
in Sect. 2 in the framework of the relativistic correct Wright
formalism. We propose a method to derive a velocity DF from
multi-frequency SZ observations in Sect. 3. Relaxation of a system
of electrons to equilibrium distributions is considered in Sect. 4.
We estimate the precision of SZ observations, which is necessary to
derive the electron DF proposed as an explanation of hard tails in
X-ray spectra of galaxy clusters, in Sect. 5. Using 3D hydrodynamic
numerical simulations of a hot merging galaxy cluster and the Wright
formalism, we demonstrate that the morphologies of the SZ intensity
maps are different at low and high frequencies in Sect. 6. We
present the observational estimates and discuss the confusion noises
in Sect. 7. We discuss our results and present our conclusions in
Sect. 8.

\section{Dependence of the CMB distortion caused by the SZ effect on the DF of electrons}

In this section we discuss the difference between the CMB
distortions (caused by the SZ effect) that are caused by a departure
from the diffusive approximation given by Kompaneets (1957) from
those that are caused by using a relativistic correct DF instead of
a Maxwell-Boltzman DF.

In the diffusion approximation and for a non-relativistic electron
population, the CMB intensity change caused by the SZ effect is

 \begin{equation}
 \Delta I_{\mathrm{nr}} =
I_{0} \tau {k_{\mathrm{b}} T_\mathrm{e} \over m_\mathrm{e}c^2}
g_{\mathrm{nr}}(x)
 \end{equation}
where  $I_{0}=2(k_{b} T_{\mathrm{CMB}})^3/(hc)^2$, $T_{\mathrm{e}}$
is the electron temperature, $\tau$ is the the scattering optical
depth, $m_{\mathrm{e}}$ the electron mass, $c$ the speed of light,
$k_{b}$ the Boltzmann constant, $x=h\nu/(k_{b} T_{\mathrm{CMB}})$,
$h$ the Planck constant, and the spectral function
$g_{\mathrm{nr}}(x)$ is given by
\begin{equation}
 g_{\mathrm{nr}}(x) = {x^4 e^{x} \over (e^{x} - 1)^2} \bigg( x \cdot {e^{x} + 1 \over e^{x} -
 1}-4\bigg).
 \end{equation}
The subscript $'nr'$ denotes that the previous expression was
obtained in the non-relativistic limit.

In the relativistic treatment taking into account the scattering to
arbitrary frequencies (Wright 1979), the CMB spectral distortion
caused by the SZ effect is a functional of the electron velocity DF
and is given by
 \begin{equation}
\Delta I\left[x, f_\mathrm{e}\right]  = I_{0} \tau
 \int ds P_1\left[s, f_\mathrm{e}\right] \bigg( {x^3 e^{-3s} \over \exp(x e^{-s})- 1} -
 {x^3 \over e^{x} - 1} \bigg)
 \label{I}
 \end{equation}
with
 \begin{equation}
 P_1\left[s, f_\mathrm{e}\right] = \int d \beta f_\mathrm{e}(\beta) P(s, \beta)
 \end{equation}
where $P_{1}\left[s, f_\mathrm{e}\right]$ is the probability
distribution of a scattering frequency shift that is a functional of
the electron velocity DF, $f_\mathrm{e}(\beta)$, of the electron
population and $P(s, \beta) ds$ is the probability that a single
scattering of a CMB photon off an electron with speed $\beta c$
causes a logarithmic frequency shift $s \equiv ln(\nu ' / \nu$), and
$\beta=V/c$, where $V$ is the electron velocity.

We notice that this formalism is valid in the single scattering
approximation and for low values of the optical depth $\tau$, which
is however sufficient for the purposes of this paper. A more general
description of the CMB spectral distortions caused by the SZ effect
can be found in Colafrancesco et al. (2003).

\begin{figure}[ht]
\centering
\includegraphics[angle=0, width=7.5cm]{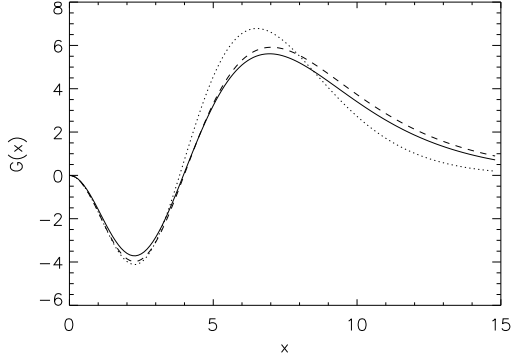}
\caption{SZ intensity spectra $G(x)=\Delta I(x)/(I_{0}y)$  for a
massive cluster with a temperature of 15.3 keV for Juttner and
Maxwell-Boltzman DFs shown by the solid and dashed lines,
respectively. The SZ intensity spectrum in the Kompaneets
approximation is shown by the dotted line.} \label{F5}
\end{figure}

The SZ intensity spectra $G(x)=\Delta I(x)/(I_{0}y)$, where $y$ is
the Comptonization parameter, derived in the framework of the Wright
formalism assuming Juttner and Maxwell-Boltzman DFs are shown in
Fig.\ref{F5} for a massive cluster with a temperature of 15.3 keV.
The SZ intensity spectrum in the Kompaneets approximation is also
shown in Fig.\ref{F5}.

\begin{figure}[ht]
\centering
\includegraphics[angle=0, width=7.5cm]{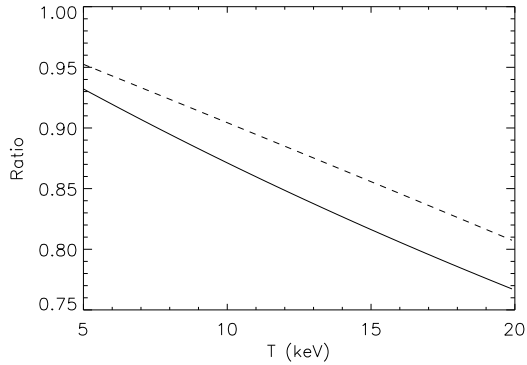}
\caption{Ratios $\Delta I\left[f_\mathrm{J}(\beta)\right]/\Delta
I_{\mathrm{nr}}$ and $\Delta
I\left[f_\mathrm{M}(\beta)\right]/\Delta I_{\mathrm{nr}}$ as
functions of the electron temperature are shown by the solid and
dashed lines, respectively}
 \label{F1}
\end{figure}

To study the effects of the CMB intensity change caused by the SZ
effect that is in turn caused by a departure from the diffusive
approximation and by using a relativistic correct DF instead of a
Maxwell-Boltzman DF, we calculate the ratios $\Delta
I\left[f_\mathrm{J}(\beta)\right]/\Delta I_{\mathrm{nr}}$ and
$\Delta I\left[f_\mathrm{M}(\beta)\right]/\Delta I_{\mathrm{nr}}$ as
functions of the electron temperature, where $f_\mathrm{J}(\beta)$
and $f_\mathrm{M}(\beta)$ are the Juttner and Maxwell-Boltzman DFs,
respectively. These ratios are shown in Fig.\ref{F1} for Juttner and
Maxwell-Boltzman DFs by solid and dashed lines, respectively, at a
frequency of 369 GHz, the choice of this frequency is reasonable
since the relativistic corrections of the SZ effect are larger at
high frequencies.

\begin{figure}[ht]
\centering
\includegraphics[angle=0, width=7.5cm]{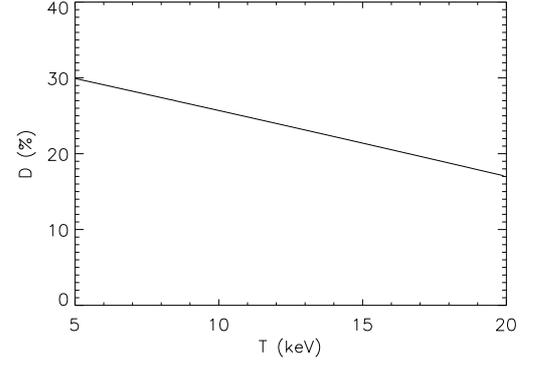}
\caption{Ratio of the CMB intensity change caused by the SZ effect
caused by using a relativistic correct DF instead of a
Maxwell-Boltzman DF to that given by the total contribution of the
SZ relativistic corrections.} \label{F2}
\end{figure}

To qualify the CMB intensity change caused by the SZ effect that is
in turn caused by using a relativistic correct DF instead of a
Maxwell-Boltzman DF, we calculate the ratio of the CMB intensity
change caused by the SZ effect that is in turn caused by using a
relativistic correct DF instead of a Maxwell-Boltzman DF to that
given by the total contribution of the SZ relativistic corrections.
This ratio is given by the expression
\begin{equation}
D=\frac{\Delta I\left[f_\mathrm{M}(\beta)\right]-\Delta
I\left[f_\mathrm{J}(\beta)\right]}{\Delta I_{\mathrm{nr}}-\Delta
I\left[f_\mathrm{J}(\beta)\right]}
\end{equation}
and is shown (in \%) at a frequency of 369 GHz in Fig.\ref{F2} as a
function of temperature.
Figure \ref{F2} shows that the correction from using a relativistic
correct DF instead of a Maxwell-Boltzman DF is a more significant
fraction of the total contribution of the SZ relativistic
corrections at lower temperatures. However, we stress that the value
of the CMB intensity change because of using a relativistic correct
DF instead of a Maxwell-Boltzman DF will be much higher in hot
galaxy clusters and, therefore, measurements of the relativistic SZ
corrections in hot clusters will be more effective to probe the
electron DF. This is because the value of the relativistic SZ
corrections is proportional to $T^{5/2}_{\mathrm{e}}$, since $\tau
\propto \sqrt{T_{\mathrm{e}}}$ (e.g., Bryan \& Norman 1998) and
\begin{equation}
\int ds P_1\left[s, f_\mathrm{e}\right] \bigg( {x^3 e^{-3s} \over
\exp(x e^{-s})- 1} - {x^3 \over e^{x} - 1} \bigg)-{k_{\mathrm{b}}
T_\mathrm{e} \over m_\mathrm{e}c^2} g_{\mathrm{nr}}(x) \propto
T^2_{\mathrm{e}},
\end{equation}
see, e.g., Challinor \& Lasenby (1998). Therefore, the relativistic
SZ corrections for a galaxy cluster with a gas temperature of 15 keV
will be in $\approx$16 times larger than that for a galaxy cluster
with a gas temperature of 5 keV. Because the second-order and
third-order (in the expansion parameter $\Theta=k_{\mathrm{b}}
T_{\mathrm{e}}/(m_{\mathrm{e}} c)$) relativistic effects make a
significant contribution to the SZ spectral distortion for
$k_{\mathrm{b}} T_{\mathrm{e}}\approx15$ keV (Challinor \& Lasenby
1998; Itoh et al. 1998), we checked the relativistic SZ corrections
for a hot galaxy cluster by using the Wright formalism and found
that the relativistic SZ corrections at frequencies of 369 GHz and
857 GHz for a galaxy cluster with a gas temperature of 15 keV are
$\approx$14 and $\approx$17 times larger than those at frequencies
of 369 GHz and 857 GHz for a galaxy cluster with a gas temperature
of 5 keV, respectively. This agrees with the result obtained above.

Because the the SZ intensity change from using a relativistic
correct DF instead of a Maxwell-Boltzman DF is a significant
fraction of the total contribution of the SZ relativistic
corrections (see Fig.\ref{F2}), we conclude that this should provide
us with the ability to derive an electron DF by means of the SZ
effect. In the following section we propose a method to derive a DF
of electrons from multi-frequency SZ observations.

\section{A method to derive a velocity DF from multi-frequency SZ observations}

We use a Fourier analysis to derive a DF of electrons from
multi-frequency SZ observations and to find the best frequencies
at which this method can be successfully applied.

\begin{figure}[ht]
\centering
\includegraphics[angle=0, width=7.5cm]{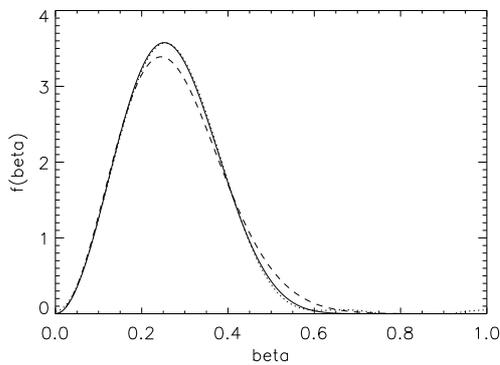}
\caption{Relativistic and non-relativistic DFs, and the
approximation to a Juttner distribution obtained by means of six
terms of Fourier series shown by the solid, dashed, and dotted
lines, respectively.} \label{F3}
\end{figure}

We decompose the Juttner and Maxwell-Boltzman functions (in the beta
representation $\beta=\mathrm{V}/c$) into the Fourier cosine series
and note that Maxwell-Boltzman and Juttner functions are
approximated with a high precision by six terms of this
decomposition (these terms are 1/2, cos($\pi\beta$),
cos($2\pi\beta$), cos($3\pi\beta$), cos($4\pi\beta$), and
cos($5\pi\beta$)). For a plasma with a temperature of 15.3 keV
(which corresponds to $\Theta=k_{\mathrm{b}}
T_{\mathrm{e}}/(m_{\mathrm{e}}c)=0.03$), the decomposition
coefficients of a Juttner function approximately equal 2.000, 1.256,
-0.172, -0.954, -0.786, -0.310 for the terms 1/2, cos($\pi\beta$),
cos($2\pi\beta$), cos($3\pi\beta$), cos($4\pi\beta$), and
cos($5\pi\beta$), respectively. For a plasma with a temperature of
15.3 keV, the decomposition coefficients of a Maxwell-Boltzman
function approximately equal 2.000, 1.214, -0.204, -0.878, -0.699,
-0.316 for the terms 1/2, cos($\pi\beta$), cos($2\pi\beta$),
cos($3\pi\beta$), cos($4\pi\beta$), and cos($5\pi\beta$),
respectively. Note that the coefficient at the first term equals 2,
because DFs are normalized. Juttner and Maxwell-Boltzman functions
for a temperature 15.3 keV are shown in Fig.\ref{F3} by solid and
dashed lines, respectively. The approximation to a Juttner
distribution obtained by means of six terms of Fourier series is
shown in Fig. \ref{F3} by a dotted line. Figure \ref{F3} shows that
six terms of the Fourier series is sufficient to describe the main
features of DFs: an increase of the Juttner DF in the range
$\beta$=[0.2, 0.4] and a decrease of the DF in the range
$\beta$=[0.4, 0.6] with respect to the Maxwell-Boltzman DF, which
arise from relativistic corrections.

To show how a velocity DF of electrons can be derived from
multi-frequency SZ observations, we write Eq.(\ref{I}) by using
the generalized spectral function $G(x, T_{\mathrm{e}})$
introduced by Prokhorov et al. (2010b) given by the expression

\begin{equation}
G(x, f_{\mathrm{e}})=\int \frac{P_1\left[s,
f_\mathrm{e}\right]}{\Theta(T_{\mathrm{e}})} \bigg( {x^3 e^{-3s}
\over \exp(x e^{-s})- 1} -
 {x^3 \over e^{x} - 1} \bigg) ds.
\end{equation}

Thus, the CMB intensity change caused by the SZ effect is $\Delta I
= I_{0} \tau \left(k_{\mathrm{b}} T_\mathrm{e}/
(m_\mathrm{e}c^2)\right)\times G(x, T_{\mathrm{e}})$. Note that in
this notation, the spectral function $g_{\mathrm{nr}}(x)$ is changed
to the generalized spectral function G(x, $T_{\mathrm{e}}$), which
explicitly depends on the electron temperature.

The CMB intensity change caused by the SZ effect for a DF
represented by six terms of Fourier cosine series is

\begin{equation}
\Delta I=I_{0} \tau \sum^{5}_{\mathrm{k}=0} A_{\mathrm{k}} \int ds
P_1\left[s, -1/2\delta_{\mathrm{k}}+\cos\left(\pi
k\beta\right)\right] K(s, x), \label{series}
\end{equation}
where $\delta_{\mathrm{k}}$ is the Kronecker delta function,
$A_{\mathrm{k}}$ are the Fourier coefficients given by

\begin{equation}
A_{\mathrm{k}}=2 \int^{1}_{0} f_{\mathrm{e}}(\beta) \cos(\pi k x) dx
\end{equation}
for k=0, 1, 2, 3, 4, 5; and

\begin{equation}
K(s, x)=\bigg({x^3 e^{-3s} \over \exp(x e^{-s})- 1} - {x^3 \over
e^{x} - 1} \bigg).
\end{equation}

Multiplying the left-hand and right-hand sides of Eq.(\ref{series})
by $1/\Theta(T_{\mathrm{e}})$, we rewrite Eq.(\ref{series}) in terms
of the generalized spectral function $G(x, T_{\mathrm{e}})$ in the
form

\begin{equation}
\frac{\Delta I}{I_{0} \tau \Theta(T_{\mathrm{e}})} =
\sum^{5}_{\mathrm{k}=0} A_{\mathrm{k}} G(x,
-1/2\delta_{\mathrm{k}}+\cos\left(\pi k\beta\right)).
\end{equation}

Since $A_{0}=2$ because of the DF normalization, we need to find
only five Fourier coefficients. Using multi-frequency measurements
of the SZ effect, we can find a solution for the linear system of
equations
\begin{equation}
\frac{\Delta I(x_{\lambda})}{I_{0} \tau \Theta(T_{\mathrm{e}})} =
\sum^{5}_{\mathrm{k}=0} A_{\mathrm{k}} G(x_{\lambda},
-1/2\delta_{\mathrm{k}}+\cos\left(\pi k\beta\right))
\label{sys}
\end{equation}
in order to derive the Fourier coefficients $A_{\mathrm{k}}$, for
$\lambda$=1, 2, 3, 4, 5.

Note that the Comptonization parameter should be independently
derived from a precise SZ intensity measurement at a frequency of
$\nu=255$ GHz (x=4.5), since the SZ intensities for Juttner and
Maxwell-Boltzman DFs have the same value at this frequency (see
Fig.\ref{F5}).

To make the system of linear equations well-conditioned, we
constrain our analysis of DFs by considering the functions for
which $f_{\mathrm{e}}(\beta=0)=0$. This choice is consistent with
both Juttner and Maxwell-Boltzman DFs. In this case, taking into
account that $A_{0}=2$, the system of linear equations (\ref{sys})
can be written as

\begin{eqnarray}
&&\sum^{5}_{\mathrm{k}=1} A_{\mathrm{k}} \left(G(x_{\lambda},
-1/2\delta_{\mathrm{k}}+\cos\left(\pi
k\beta\right))-g_{\mathrm{nr}}(x_{\lambda})\right)=\\ \nonumber
&&=g_{\mathrm{nr}}(x_{\lambda})+ \frac{\Delta I(x_{\lambda})}{I_{0}
\tau \Theta(T_{\mathrm{e}})}-\int\frac{P_{1}[s,
1]}{\Theta(T_{\mathrm{e}})} K(s, x_{\lambda}) ds.
\end{eqnarray}

To check if this system of linear equations is well-conditioned, we
study the properties of the matrix $M_{\mathrm{\lambda k}}$,  which
is given by
\begin{equation}
M_{\mathrm{\lambda} k}=G(x_{\lambda},
-1/2\delta_{\mathrm{k}}+\cos\left(\pi
k\beta\right))-g_{\mathrm{nr}}(x_{\lambda})
\end{equation}
for different samples of frequencies.

To find a suitable sample of frequencies for deriving the electron
DF, let us consider the following samples: the first sample is in
the low-frequency range including four frequencies $\nu=$ 100, 120,
140, 160 GHz considered by Colafrancesco \& Marchegiani (2010); the
second sample including four frequencies in the frequency range
300-400 GHz considered by Colafrancesco \& Marchegiani (2010), $\nu$
= 300, 320, 340, 360 GHz; the third sample including low and high
frequencies $\nu$ = 100, 200, 300, 400 GHz; and the fourth sample
including more higher frequencies
$\nu$ = 375, 600, 700, 857 GHz.\\
The first detection of the Sunyaev-Zel'dovich effect increment at
such high frequencies where the relativistic corrections of the SZ
effect are relevant has been obtained with HERSCHEL-SPIRE for the
Bullet cluster (Zemcov et al. 2010).

To find the frequency sample for which the electron DF can be
optimally derived, we calculate the condition numbers for the
matrixes of $M_{\mathrm{\lambda} k}$ and the inverse matrixes of
$M^{-1}_{\mathrm{k \lambda}}$. Since these samples have only four
frequencies, the equation obtained from the condition of
$f_{\mathrm{e}}(\beta=0)=0$ is used $A_{1}+A_{2}+A_{3}+A_4+A_5=-1$
(the normalization condition $A_{0}=2$ is taken into account) to
form the matrix $5\times5$, we choose the first row of these
matrixes equaled to $M_{1 k}=(1, 1, 1, 1, 1)$. The condition numbers
based on the $L_{2}$ norm of these matrixes approximately equal to
$4.4\times10^4$, $2.1\times10^5$, $1.6\times10^3$, and 323.8 for the
1st, 2nd, 3rd, and 4th samples, respectively. Since the condition
number of the matrix corresponding to the fourth frequency sample is
the smallest, we conclude that the best sample to determine DFs by
means of the SZ effect is the fourth sample, which includes high
frequencies. The worst sample is the second sample, which is in the
frequency range 300-400 GHz, since the matrix $M_{\mathrm{\lambda k}
(2)}$ is ill-conditioned.

To clarify our method of the DF analysis, we calculate the inverse
matrixes

\begin{equation}
M^{-1}_{\mathrm{k \lambda} (1)}=\left(\begin{array}{ccccc}
-0.86 & 0.15 & -0.59 & 1.59 & -1.05\\
-3.11 & -19.35 & 18.60 & 5.30 & -6.74\\
-5.28 & 190.19 & -522.48 & 499.98 & -164.28\\
2.00 & 159.65 & -358.39 & 294.83 & -89.09\\
8.25 & -330.66 & 862.87 & -801.71 & 261.18
\end{array}\right)
\end{equation}

\begin{equation}
M^{-1}_{\mathrm{k \lambda} (2)}=\left(\begin{array}{ccccc}
-0.89 & -6.89 & 25.02 & -28.79 & 10.69\\
-3.83 & 42.01 & -112.45 & 97.69 & -27.23\\
-6.68 & 808.77 & -2462.02 & 2509.77 & -856.15\\
2.79 & 331.94 & -1046.14 & 1104.08 & -390.21\\
9.63 & -1175.84 & 3595.60 & -3682.75 & 1262.90
\end{array}\right)
\end{equation}

\begin{equation}
M^{-1}_{\mathrm{k \lambda} (3)}=\left(\begin{array}{ccccc}
-0.88 & 0.28 & -0.14 & -0.03 & 0.02\\
-3.75 & -0.90& 1.76 & -2.14 & 1.01\\
-6.76 & -22.04 & 25.24 & -11.68 & 1.06\\
2.57 & -9.60 & -1.86 & -1.86 & -1.71\\
9.82 & 32.25 & -36.24& 15.70 & -0.38
\end{array}\right)
\end{equation}

\begin{equation}
M^{-1}_{\mathrm{k \lambda} (4)}=\left(\begin{array}{ccccc}
-0.21 & 0.31 & -2.18 & 2.79 & -0.96\\
-2.36 & -0.06 & -2.43 & 3.95 & -1.61\\
-2.79 & -0.38 & -2.06 & 5.24 & -3.19\\
0.86 & -1.00 & 6.94 & -9.18 & 2.95\\
5.49 & 1.12 & -0.28 & -2.79 & 2.81
\end{array}\right)
\end{equation}

The experimental uncertainties in the SZ intensity measurements
constrain the ability to derive DFs of electrons. The uncertainties
in the determination of the Fourier coefficients of $A_{k}$ are
given by

\begin{equation}
\delta {A}_{j}= \sum_{\lambda} M^{-1}_{\mathrm{j \lambda}}
\frac{\delta\left(\Delta I(x_{\lambda})/I_{0}\right)}{\tau
\Theta(T_{\mathrm{e}})}
\end{equation}
and can be written as

\begin{equation}
\delta {A}_{j}=  \sum_{n} M^{-1}_{\mathrm{j n}} \sum_{\lambda}
\tilde{G}_{\mathrm{n} \lambda} V_{\lambda} \label{A},
\end{equation}
where
\begin{equation}
V_{\lambda}=\left(\frac{\delta\left(\Delta I(x)\right)}{\Delta
I(x)}\right)_{x=x_{\lambda}}
\end{equation}
for $\lambda=2, 3, 4, 5$ and $V_{1}=0$ (since the first term
determined from the condition $f(\beta=0)=0$), and

\begin{equation}
\tilde{G}_{\mathrm{n} \lambda}=\left(\begin{array}{ccccc}
0 & 0 & 0 & 0 & 0\\
0 & G(x_{1}, T_{\mathrm{e}}) & 0 & 0 & 0\\
0 & 0 & G(x_{2}, T_{\mathrm{e}}) & 0 & 0\\
0 & 0 & 0 & G(x_{3}, T_{\mathrm{e}}) & 0\\
0 & 0 & 0 & 0 & G(x_{4}, T_{\mathrm{e}})
\end{array}\right)
\end{equation}
where $x_{1}, x_{2}, x_{3}, x_{4}$ correspond to the considered
frequencies ($\nu$ = 375, 600, 700, 857 GHz for the fourth sample)
and the generalized spectral functions $G(x_{\lambda},
T_{\mathrm{e}})$ are calculated for a Juttner DF (this is because we
study the ability of multi-frequency SZ observations to distinguish
a Juttner DF from other DFs, which are slightly different from a
Juttner DF).

Equation (\ref{A}) can be written as
\begin{equation}
\delta {A}_{j}= \sum_{\lambda} W_{\mathrm{j} \lambda} V_{\lambda}
\label{V}
\end{equation}
where
\begin{equation}
W_{\mathrm{j} \lambda}=\sum_{n} M^{-1}_{\mathrm{j n}}
\tilde{G}_{\mathrm{n} \lambda}.
\end{equation}

The matrixes $W_{\mathrm{j} \lambda}$ are shown below for the
samples of frequencies $\nu$ = 375, 600, 700, 857 GHz and $\nu$ =
300, 320, 340, 360 GHz for a comparison.

\begin{equation}
W_{\mathrm{j} \lambda (4)}=\left(\begin{array}{ccccc}
0 & 1.75 & -6.27 & 4.61 & -0.62\\
0 & -0.33 & -6.99 & 6.53 & -1.03\\
0 & -2.10 & -5.92 & 8.67 & -2.04\\
0 & -5.54 & 19.99 & -15.19 & 1.89\\
0 & 6.22 & -0.80 & -4.62 & 1.80
\end{array}\right)
\label{W}
\end{equation}

\begin{equation}
W_{\mathrm{j} \lambda (2)}=\left(\begin{array}{ccccc}
0 & -27.07 & 115.26 & -146.87 & 57.94\\
0 & 165.02 & -517.97 & 498.23 & -147.59\\
0 & 3176.87 & -11340.09 & 12799.83 & -4640.34\\
0 & 1303.88 & 4818.54 & 5630.83 & -2114.94\\
0 & -4618.70 & 16561.34 & -18782.03 & 6844.92
\end{array}\right)
\end{equation}

Since the absolute values of the elements of $\tilde{W}_{\mathrm{j}
\lambda (4)}$ are much smaller than those of $\tilde{W}_{\mathrm{j}
\lambda (2)}$, we justify our previous conclusion that the fourth
sample is more suitable for the SZ analysis of electron DF than the
second sample, since in the case of the fourth sample to derive a DF
of electrons the allowable uncertainties in the SZ intensities are
about three orders of magnitude larger (see Eq.\ref{V}). We also
checked that the fourth sample is the best amongst the all samples
considered in this paper in order to study DFs.

To study the ability to derive DFs from SZ intensity measurements,
we calculate the differences between the corresponding Fourier
coefficients for Juttner and Maxwell-Boltzman DFs, which are $|A_{1,
\mathrm{J}}-A_{1, \mathrm{MB}}|=0.042$, $|A_{2, \mathrm{J}}-A_{2,
\mathrm{MB}}|=0.032$, $|A_{3, \mathrm{J}}-A_{3,
\mathrm{MB}}|=0.076$, $|A_{4, \mathrm{J}}-A_{4,
\mathrm{MB}}|=0.087$, and $|A_{5, \mathrm{J}}-A_{5,
\mathrm{MB}}|=0.006$ (see the previous section).

Note that the difference $|A_{5, \mathrm{J}}-A_{5,
\mathrm{MB}}|=0.006$ is very small and it will be difficult to
observe this difference by the SZ measurements. However, we checked
that this difference has a negligible impact on the precision of the
approximation of a Juttner DF obtained by means of Fourier series
because of its small value.

For the sake of illustration, we assume that the relative SZ
intensity uncertainty in measurements ${\delta\left(\Delta
I(x_{\lambda})\right)}/{\Delta I_{x_{\lambda}}}$ does not depend
on the frequency and equals $\xi$.

From Eqs.(\ref{V}) and (\ref{W}) we find that the uncertainty in SZ
intensity measurements to distinguish the values of the Fourier
coefficients for Juttner and Maxwell-Boltzman DFs should be less
than $\xi=0.32\%$ to derive the $A_{1}$ value, $\xi=0.22\%$ to
derive the $A_{2}$ value, $\xi=0.41\%$ to derive the $A_{3}$ value,
and $\xi=0.20\%$ to derive the $A_{4}$ value.

\begin{figure}[ht]
\centering
\includegraphics[angle=0, width=7.5cm]{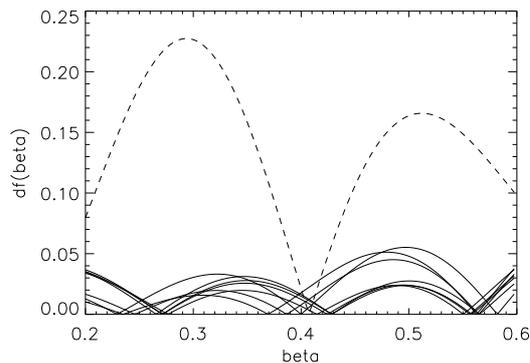}
\caption{Absolute difference between the Juttner and approximate DFs
for a temperature of 15.3 keV (solid lines) compared with the
absolute difference between the Juttner and Maxwell-Boltzman DFs
(dashed line).} \label{F4}
\end{figure}

To check the result that an electron DF can be derived by means of
the SZ effect if uncertainties of SZ observational data are less
than 0.2 \%, we performed Monte-Carlo simulations of SZ observations
(calculated in the relativistic correct formalism and using a
Juttner DF) with the SZ intensity uncertainty of 0.1\% at the
frequencies included in the fourth sample. Using the SZ spectra
obtained from the Monte-Carlo simulations, we solve the system of
equations (13) to find the Fourier coefficients $A_{\mathrm{k}}$ and
the approximate functions to a Juttner DF, which has been initially
used to produce the CMB intensity changes caused by the SZ effect in
Monte-Carlo simulations. The absolute value of the difference
between the Juttner and Maxwell-Boltzman DFs for a temperature of
15.3 keV in the range of $\beta=[0.2, 0.6]$ is shown in Fig.\ref{F4}
by a dashed line. This range of $\beta$ contains the main features
of these DFs: an increase of the Juttner DF in the range
$\beta=[0.2, 0.4]$ and a decrease of the Juttner DF in the range
$\beta=[0.4, 0.6]$ with respect to the Maxwell-Boltzman DF. The
absolute difference between the Juttner DF for a temperature of 15.3
keV and the approximate functions to the Juttner DF obtained by the
Monte-Carlo simulations are shown in Fig.\ref{F4} by solid lines.
Figure \ref{F4} shows that the approximate DFs are close to a
Juttner DF and it is possible to distinguish the Juttner and
Maxwell-Boltzman DFs, if the SZ intensities are measured with an
uncertainty of 0.1\% (see Sect.6 below).

We conclude that the SZ effect provides us with an interesting
method to study electron DFs in massive galaxy clusters that contain
hot plasmas with temperatures ($\simeq 15$ keV) .

\section{Relaxation of a system of electrons to equilibrium distributions}

To justify the use of a Juttner distribution function as an
appropriate approximation to the universal (equilibrium)  electron
distribution in hot merging clusters, we calculate the electron
equilibration time and compare this time with the merging time
scale.
We consider here the relaxation to equilibrium distributions in the
framework of both the non-relativistic and relativistic theories.

We solve numerically the time-dependent Fokker-Planck equation for
the evolution of an isotropic system of electrons with Coulomb
interactions and with an initially Gaussian momentum distribution.
The Fokker-Planck equation coefficients are taken from Liang et al.
(2002) and Dogiel et al. (2007). The initial Gaussian momentum
distributions are chosen so that the number density normalization
and the total energy are consistent with those of the equilibrium
distributions. Note that both the Fokker-Planck coefficients (see
Dogiel et al. 2007) are proportional to $m_{\mathrm{e}}
c^2/(k_{\mathrm{b}} T_{\mathrm{e}})$ and, therefore, the value of
the characteristic frequency (which determines the rate of
relaxation) is given by the inverse Spitzer collision time (see Eq.
18 of Liang et al. 2002). We use here the dimensionless time $\tau$
in units of the Spitzer collision time.
For an electron population with number density of $n=10^{-3}$
cm$^{-3}$ and temperature $T_{\mathrm{e}}=15.3$ keV, the Spitzer
collision time equals $1.3\times10^5$ yr.

\begin{figure}[ht]
\centering
\includegraphics[angle=0, width=7.5cm]{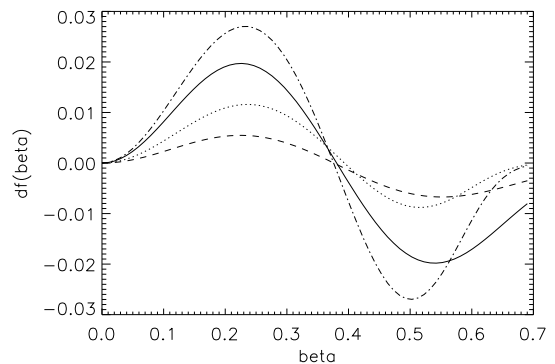}
\caption{Difference between the electron distribution function and
the Maxwell-Boltzman DF at time $\tau=40$ (solid) and $60$ (dashed)
is shown as a function of $\beta=v/c$. The analogous difference
between the electron distribution function and the Juttner  DF at
time $\tau=40$ (dash-dotted) and $60$ (dotted) is also shown.}
 \label{relax}
\end{figure}

The results of our numerical analysis of relaxation of a system of
electrons to equilibrium distributions are plotted in Fig.
\ref{relax}.

The solid and dashed curves in Fig.\ref{relax} show the difference
between the electron distribution function, derived in the framework
of the non-relativistic theory, at times $\tau=40$ and $60$,
respectively, and the reference Maxwell-Boltzman DF. Since these
differences are significantly smaller than the difference between
the Juttner and Maxwell-Boltzman DFs shown in Fig. \ref{F4}, the
electron DFs at time $\tau=40$ and $60$ can be considered as good
approximations to the Maxwell-Boltzman DF.

The dash-dotted and dotted curves in Fig.\ref{relax} show the
difference between the electron distribution function, derived in
the framework of the relativistic theory, at times $\tau=40$ and
$60$, respectively, and the reference Juttner DF. These differences
are again significantly smaller than the difference between the
Juttner and Maxwell-Boltzman DFs shown in Fig. \ref{F4} and,
therefore, the electron DFs at a time of $\tau=40$ and $60$ can be
considered as good approximation to the Juttner DF.

We stress that the qualitative estimate of the time required to fill
the Maxwell tail to the velocity of $0.7c$ in the non-relativistic
theory (see, e.g., MacDonald \& Rosenbluth 1957) is twenty five
Spitzer collision times and is in rough agreement with our numerical
analysis.

Since sixty Spitzer collision times correspond to $\approx
8.1\times10^6$ yrs for the number density and temperature values
mentioned above, this time scale is a tiny fraction of the Hubble
time. The merging time scale of a merging cluster equaled to $\simeq
10^{8}-10^{9}$ yrs (see, e.g., Prokhorov \& Durret 2007; Akahori \&
Yoshikawa 2010) is  longer than sixty Spitzer collision times. Thus,
we conclude that the Juttner distribution function is an appropriate
approximation to the universal (equilibrium) electron distribution
in hot merged clusters.

We estimate the gas density at which the time-dependent DF relaxing
to a Maxwell-Boltzman DF is indistinguishable from a Juttner DF. The
time, at which the difference between the time-dependent DF and the
Maxwell-Boltzman DF is $\approx$0.2 and at which the time-dependent
DF is close to a Juttner DF (see Fig.\ref{F4}), corresponds to five
Spitzer times or $\simeq 6.5\times 10^5$ yrs for an electron number
density of $10^{-3}$ cm$^{-3}$. Using a merging time scale of
$\gtrsim 10^8$ yrs, we found that the value of the critical number
density $\lesssim 6.5\times10^{-6}$ cm$^{-3}$. This electron number
density is much smaller than the average gas density in galaxy
clusters (see e.g. Sarazin 1986) and, therefore, it is possible to
distinguish a Juttner DF from a Maxwell-Boltzman DF in galaxy
clusters by the method proposed in the Sect. 3.

\section{DFs in galaxy clusters with hard X-ray spectral tails}

Galaxy clusters with a hard X-ray excess are promising targets to
test electron DFs by means of multi-frequency SZ observations. This
is because these clusters show evidence of either a very high
temperature plasma or a quasi-thermal emission tail due to MHD
acceleration mechanisms in the cluster atmosphere.

Petrosian (2001) estimated the yield in non-thermal bremsstrahlung
photons and found that a large amount of the energy of the
non-thermal electrons is transferred to the background plasma, so
that the ICM should be heated to above its observable temperature
within 10 Myr. However, it was shown by Liang et al. (2002) and
Dogiel et al. (2007) that a quasi-thermal electron population might
overcome this difficulty by means of a higher radiative efficiency
(and therefore a longer overheating time, but see Petrosian \& East
2008). Wolfe \& Melia (2008) also considered a quasi-thermal
electron distribution when fitting hard X-ray emission, but rather
than requiring a second-order Fermi acceleration to produce
quasi-thermal electrons, they assumed that quasi-thermal electrons
are produced by collisions with non-thermal protons.

A study of the influence of suprathermal electrons on the SZ effect
was made for the Coma and Abell 2199 clusters by Blasi et al.
(2000), Shimon \& Rephaeli (2002), and Colafrancesco et al. (2009).
An alternative probe to study the electron distribution in galaxy
clusters, namely the flux ratio of the emission lines caused by
FeK$\alpha$ transitions (FeXXV and FeXXVI) was proposed by Prokhorov
et al. (2009). This flux ratio is very sensitive to the population
of electrons with energies higher than the ionization potential of a
FeXXV ion (which is 8.8 keV). Kaastra et al. (2009) demonstrated
that the relative intensities of the satellite lines are sensitive
to the presence of suprathermal electrons in galaxy clusters.
Prokhorov (2009) studies the influence of high-energy electron
populations on metal abundance estimates in galaxy clusters and
shows that the effect of high-energy particles can be significant.
However, the mentioned approaches do not allow us to derive electron
DFs in galaxy clusters with a hard X-ray excess.

In the section, we propose an approach to derive the electron DF and
demonstrate that multi-frequency SZ observations are promising for
this purpose.

\begin{figure}[ht]
\centering
\includegraphics[angle=0, width=7.5cm]{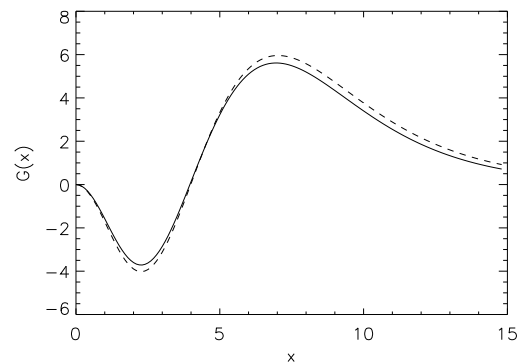}
\caption{SZ intensity spectra $G(x)=\Delta I(x)/(I_{0}y)$ for the
cluster with the hard X-ray tail for DFs with and without the second
electron population shown by the dashed and solid lines,
respectively.} \label{F6}
\end{figure}

Evidence for non-thermal X-ray emission from the Bullet cluster
reported by Petrosian et al. (2006) and Million \& Allen (2009)
suggests a possible high-energy subrelativistic electron component
if these X-ray spectra are interpreted in terms of bremsstrahlung
emission. Therefore, the Bullet cluster is an interesting target to
test electron DFs. We calculate the SZ intensity spectra for the
Bullet cluster for DFs with and without the second electron
population proposed by Petrosian et al. (2006). These SZ spectra are
shown in Fig.\ref{F6}. Here we assume that the second thermal
electron population with $k_{\mathrm{b}} T_{\mathrm{e}, 2}=50$ keV
mimics the presence of quasi-thermal electrons in the DF and that
the second electron population is 5\% of the first thermal electron
population. This fraction of high-energy subrelativistic electrons
is consistent with that derived by Petrosian et al. (2006). For the
sake of comparison with the results obtained in the previous
section, we assume that the temperature of the first thermal
component equals 15.3 keV, which agrees with the temperature values
of 14.5$^{+2.0}_{-1.7}$ keV and 14.8$^{+1.7}_{-1.2}$ keV derived by
using both ASCA and ROSAT (Liang et al. 2000) and by using Chandra
data (Markevitch et al. 2002), respectively. Note that the total DF
is normalized. The Comptonization parameter should be independently
derived from a precise SZ intensity measurement at a frequency
$\nu=250$ GHz (x=4.4), since the SZ intensities have the same value
at this frequency for normalized Juttner DFs with and without the
second electron population (see Fig.\ref{F6}).

\begin{figure}[ht]
\centering
\includegraphics[angle=0, width=7.5cm]{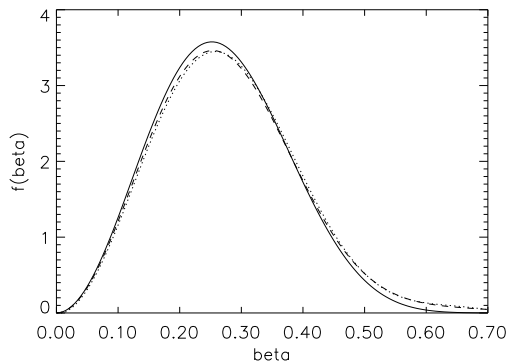}
\caption{Normalized Juttner DFs with and without the second electron
population, and the approximation to a Juttner distribution with the
second electron population obtained by means of the simulated SZ
observations shown by the solid, dashed, and dotted lines,
respectively.} \label{F7}
\end{figure}

For a plasma with these parameters, the decomposition coefficients
of the DF with the second electron component approximately equal to
2.000, 1.207, -0.220, -0.917, -0.741, -0.298 for the terms 1/2,
cos($\pi\beta$), cos($2\pi\beta$), cos(3$\pi\beta$),
cos(4$\pi\beta$), and cos(5$\pi\beta$), respectively.
To analyze the DF, we use the method of multi-frequency SZ
observations at the frequencies of $\nu$=375, 600, 700, 857 GHz
described in the previous section. To study our ability to derive
DFs from SZ intensity measurements, we calculate the differences
between the corresponding Fourier coefficients for normalized
Juttner DFs with and without the second electron component, which
are $|\Delta A_{1}|=0.049$, $|\Delta A_{2}|=0.047$, $|\Delta
A_{3}|=0.037$, $|\Delta A_{4}|=0.045$, and $|\Delta A_{5}|=0.012$.
Note that the difference $|\Delta A_{5}|=0.012$ is small and it will
be difficult to observe this difference by the SZ measurements.
However, we have checked that this difference has a negligible
impact on the precision of the approximation of the DF obtained by
means of Fourier series because of its small values.\\
From Eqs.(\ref{V}) and (\ref{W}) we find that the uncertainty in SZ
intensity measurements to distinguish the values of the Fourier
coefficients for normalized Juttner DFs with and without the second
electron component should be less than: $\xi$ = 0.37\% to derive the
A$_1$ value, $\xi$ = 0.32\% to derive the A$_2$ value, $\xi$ =
0.20\% to derive the A$_3$ value, and $\xi$ = 0.1\% to derive the
A$_4$ value. Therefore, the electron DF can be derived by means of
the SZ effect if uncertainties of SZ
observational data are less than 0.1\%.\\
To show that multi-frequency SZ observations allow us to derive
the Juttner DF with the second electron component, we calculate
the SZ intensity (Eq.\ref{I}) at the frequencies of $\nu$=375,
600, 700, 857 GHz assuming this DF. We find the decomposition
coefficients from Eq.(13) by means of the simulated SZ
observations and show the obtained approximation to the Juttner
distribution with the second electron population in Fig.\ref{F7}
by a dotted line in the beta range of $\beta=[0, 0.7]$. The
normalized Juttner DFs with and without the second electron
population are also shown in Fig.{\ref{F7}} by dashed and solid
lines, respectively.

Since the approximate function obtained by means of the simulated SZ
observations coincides precisely with the normalized Juttner DF with
the second electron population, we conclude that precise
observations of the SZ effect can allow to derive the electron DFs
in galaxy clusters with a hard X-ray excess if electron DFs are
different from a Juttner DF.

\section{SZ intensity maps for the simulated merging galaxy cluster}

In the previous sections, we showed that high-frequency spectral SZ
observations at frequencies $\nu$ = 375 GHz, 600 GHz, 700 GHz, and
857 GHz provide us with a method to derive DFs of electrons in
galaxy clusters.

Here we study the intensity maps of hot clusters at these
frequencies. Note that all the SZ intensity maps derived in the
Kompaneets approximation have the same spatial morphology. Since the
shape of the SZ effect is most sensitive to an electron DF at
frequencies $\nu$ = 375 GHz, 600 GHz, 700 GHz, and 857 GHz (see
Sects. 3 and 4), the spatial SZ intensity maps, which are derived in
the Wright formalism, at these frequencies should have different
spatial morphology.

In this section, we show that the spatial morphologies of the SZ
intensity maps are different at frequencies $\nu$ = 375 GHz, 600
GHz, 700 GHz, and 857 GHz. To this aim we use the 3D numerical
hydrodynamic simulations of a merging hot galaxy cluster presented
in Akahori \& Yoshikawa (2010). These authors considered a head-on
encounter of two free-falling galaxy clusters from the turn-around
radius (for a review, see Sarazin 2002). We assume in the simulation
that the two galaxy clusters have equal virial mass of
$M_{\mathrm{vir}}=8\times10^{14}$ M$_{\odot}$, and the impact
parameter is zero (for details on the simulation, see Akahori \&
Yoshikawa 2010).

\begin{figure}[ht]
\centering
\includegraphics[angle=0, width=7.5cm]{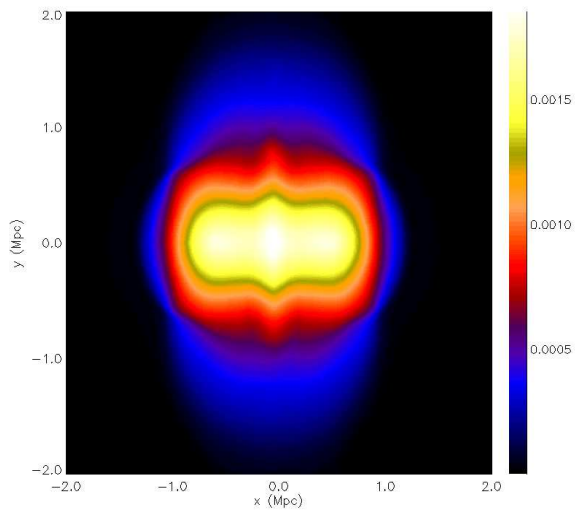}
\caption{Intensity map $I/I_0$ of the SZ effect at a frequency 375
GHz derived from the numerical simulation in the framework of the
Wright formalism.} \label{F8}
\end{figure}

\begin{figure}[ht]
\centering
\includegraphics[angle=0, width=7.5cm]{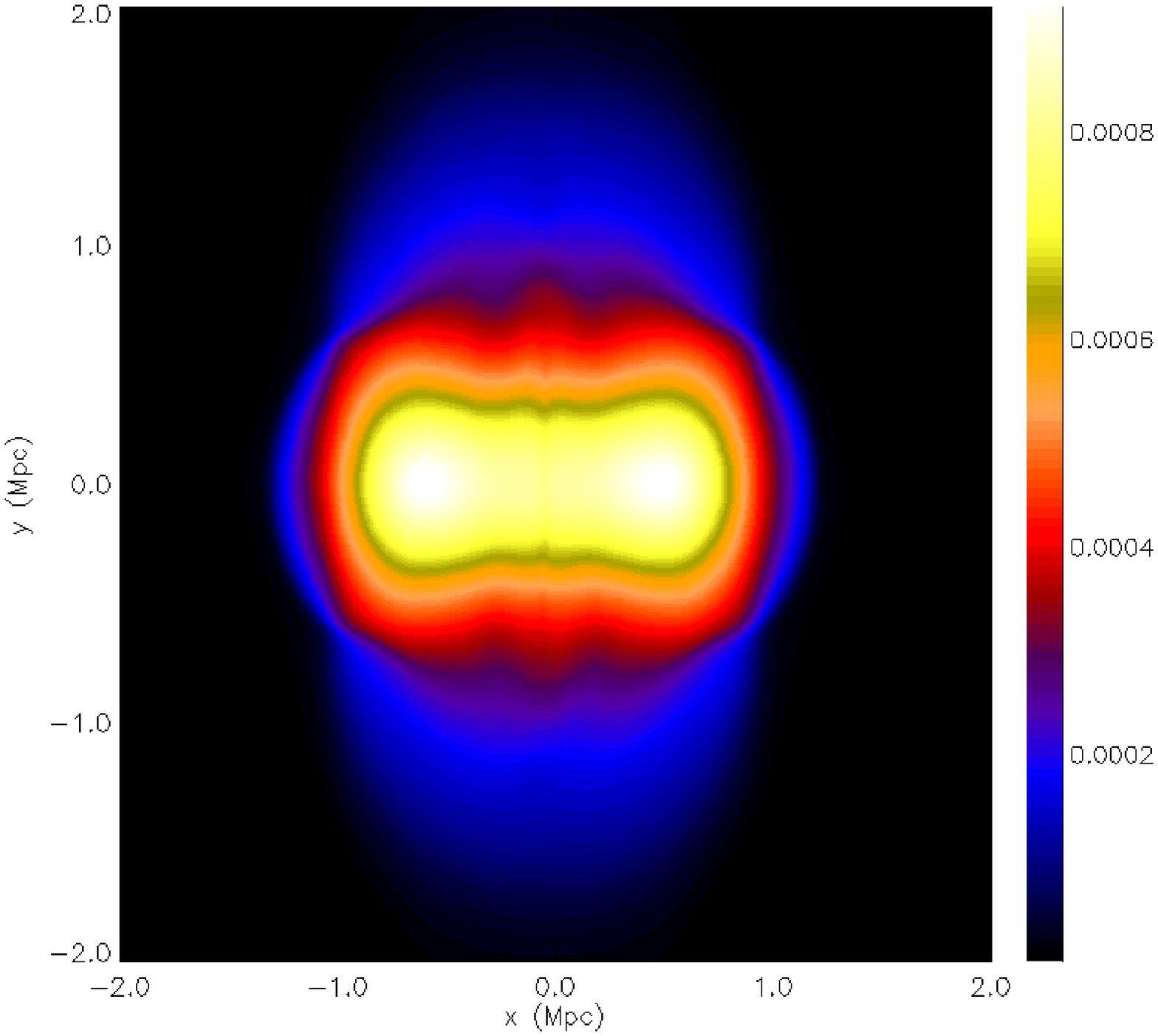}
\caption{Same as Fig.\ref{F8} but for a frequency of 600 GHz. }
\label{F9}
\end{figure}

\begin{figure}[ht]
\centering
\includegraphics[angle=0, width=7.5cm]{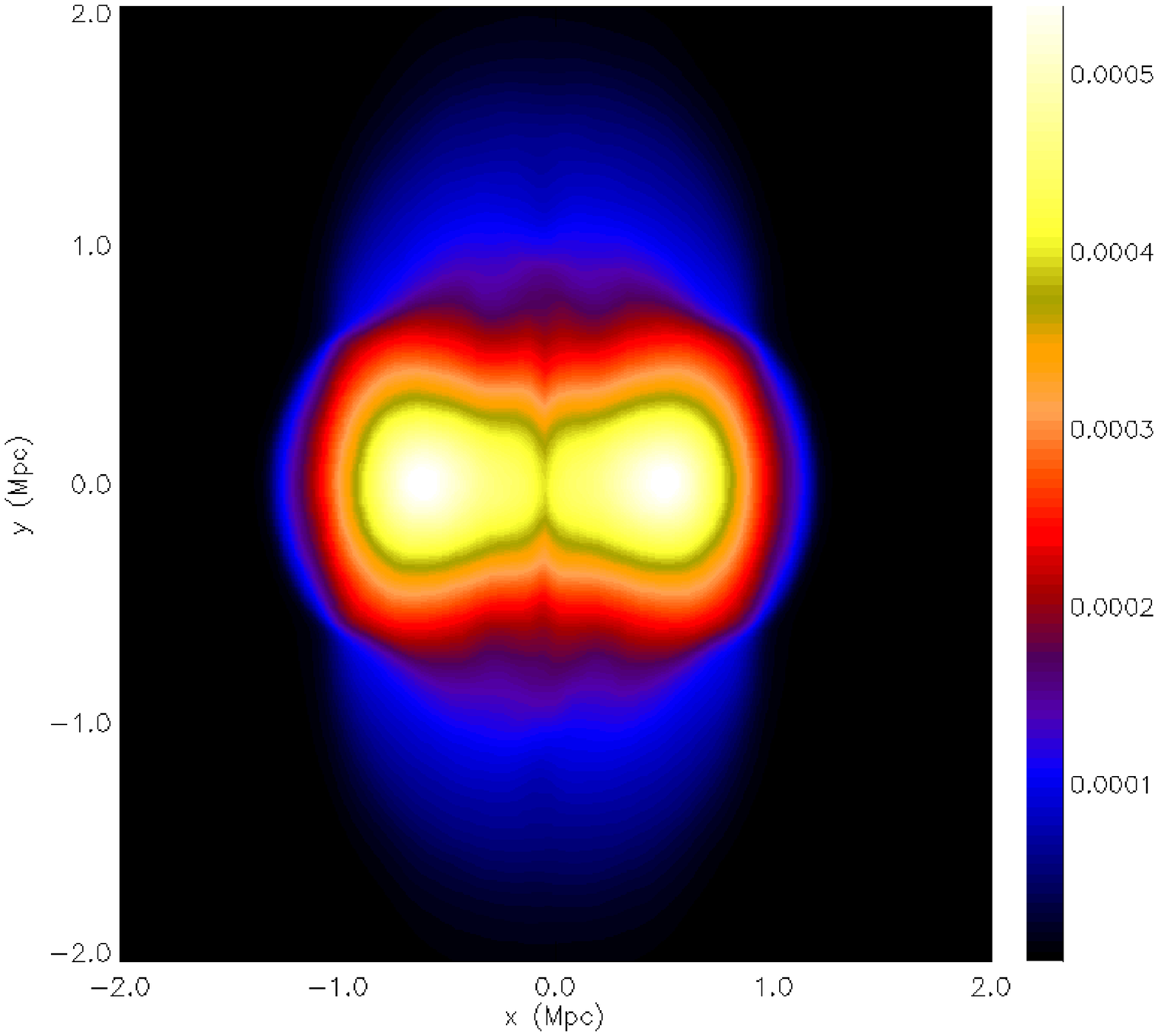}
\caption{Same as Fig.\ref{F8} but for a frequency of 700 GHz.
}\label{F10}
\end{figure}

\begin{figure}[ht]
\centering
\includegraphics[angle=0, width=7.5cm]{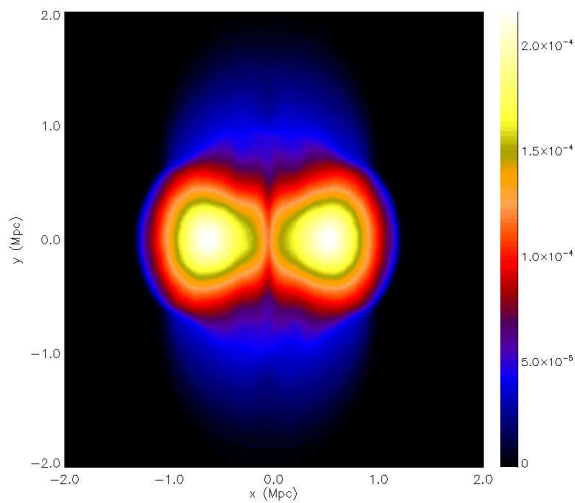}
\caption{Same as Fig.\ref{F8} but for a frequency of 857
GHz.}\label{F11}
\end{figure}

To produce the SZ intensity maps at frequencies $\nu$ = 375 GHz,
600 GHz, 700 GHz, and 857 GHz we use the 3D density and
temperature maps (see Figs. 2 and 3 from Akahori \& Yoshikawa
2010) for the simulated merging galaxy cluster at a time of t=0.5
Gyr, where t=0 Gyr corresponds to the time of the closest approach
of the centers of the dark matter halos. We calculated the SZ
intensity maps using the Wright formalism and the approach
described in Prokhorov et al. (2010a, 2010b). The intensity maps
of the SZ effect at these frequencies derived from the simulated
maps of the gas density and temperature are shown in Figs.
\ref{F8}, \ref{F9}, \ref{F10}, and \ref{F11}, respectively.

The morphologies of the SZ intensity simulated maps at  $\nu$ = 375
GHz, 600 GHz, 700 GHz, and 857 GHz are clearly different. This is
because the SZ effect from hot galaxy clusters at high frequencies
is sensitive to the relativistic effects in their electron DF and
cannot be described in the framework of the Kompaneets
approximation. Note that the morphologies of the SZ intensity
simulated maps at frequencies $\nu$ = 128 GHz and 369 GHz in
Prokhorov et al. (2010a) are instead more similar because the SZ
effect from cool galaxy clusters is less sensitive to the
relativistic effects.

We find that the SZ intensity simulated map at
frequency of 375 GHz is similar to the SZ intensity simulated map
derived in the Kompaneets approximation, while the SZ intensity
simulated map at a frequency of 857 GHz is similar to the SZ
intensity simulated map at a frequency of 217 GHz where the SZ
effect in the framework of the Kompaneets approximation is zero.
The contribution of relativistic corrections to the SZ signal at
frequency of 857 GHz dominates over that derived in the Kompaneets
approximation.

We find that the maximum of the SZ intensity increment at a
frequency of 375 GHz (see Fig. \ref{F8}) is at the center of this
map, while the maximum of the SZ intensity increment at a frequency
of 857 GHz (see Fig. \ref{F11}) is in the post-shock regions (see
the Mach number distribution of ICM in Fig. 2 of Akahori \&
Yoshikawa 2010). This is because the gas temperature in the
post-shock regions is the highest and SZ relativistic corrections
are therefore the most significant in these regions.

Comparing Fig. \ref{F8} with Fig. \ref{F11} shows that the SZ
intensity maps at low and high frequencies look different for a hot
merging galaxy cluster and, therefore, multi-frequency SZ
observations with a high spatial resolution are necessary to
demonstrate the different morphologies of SZ intensity maps at low
and high frequencies. We conclude that measurements of the SZ
intensity maps at frequencies $\nu$ = 375 GHz, 600 GHz, 700 GHz, and
857 GHz are relevant to study the importance of SZ effect
relativistic corrections.

Using the X-ray observations of the Bullet cluster and Abell 2219
(Million \& Allen 2009), which are massive merging galaxy clusters,
we calculated the SZ intensity maps at low and high frequencies and
have checked that the morphologies of the SZ intensity maps at low
and high frequencies are different. We will address this problem
more specifically in a forthcoming paper.

\section{Observational considerations}

We briefly discuss here the detectability of the electron DF through
SZ effect observations in the optimal frequency sample
studied above.\\
We find that the difference of the SZ effect signals predicted for
the Juttner and the Maxwell-Boltzmann DFs at 375, 600, 700 and 857
GHz can be only marginally appreciated by the PLANCK-HFI instrument
with the nominal 2-year survey sensitivity (at 1 $\sigma$ level).
Indeed, while the DF difference in the predicted SZ effect
intensities can be determined at 375 and 600 GHz channels at the
$\sim 3.5 \sigma$ and $\sim 3.4 \sigma$ level, it is impossible to
determine it for the highest channels, i.e. 700 and 857 GHz of
PLANCK-HFI. Note also that the Bullet cluster is
almost unresolved for the PLANCK-HFI frequency channels.\\
However, the 600 and 857 GHz frequency channels are also covered by
the HERSCHEL-SPIRE instrument, whose sensitivity is sufficient to
detect the predicted SZ effect intensity difference necessary to
distinguish between the Juttner and Maxwell-Boltzmann DFs at 600 and
857 GHz. We also notice that the Bullet cluster is a fully
resolved source for the HERSCHEL-SPIRE instrument.\\
An important point to be addressed is the impact of the confusion
noises at these high-frequency channels to distinguish the DF
effects in the SZ effect observations. Relevant sources of confusion
for SZ effect observations at high frequencies are the CMB
fluctuation, the unresolved point-like millimeter sources emission,
and the diffuse emission from the Galaxy. For an experiment that can
spatially resolve the Bullet cluster, the CMB confusion is
negligible on sub-arcmin scales, while the unresolved point-like
source emission and the Galaxy emission increase with frequency,
which provides the major confusion sources. For cluster temperatures
on the order of 15 keV (as in the Bullet cluster),
the diffuse Galaxy emission is the dominant confusion noise.\\
Therefore, a multi-frequency observational strategy is required to
properly monitor and subtract the confusion noises that have a
different frequency behavior with respect to the SZ effect spectrum.
Nonetheless, the confusion noise level adds up to the instrument
noise to reduce the possibility of the SZ effect observations in
order to derive the electron DF. Longer exposure with spectroscopic
instruments operating in this high-frequency range are required  to
reach a precision level on the order of $\sim 0.1 \%$ to achieve a
good statistical control of the systematics and a good statistical
confidence level of the parameters necessary to reconstruct the
electron DF.

\section{Conclusions}

The SZ effect is an important tool for cosmology and for the
astrophysical study of clusters of galaxies (for a review, see
Birkinshaw 1999). It measures the pressure of an electron population
integrated along the line of sight as long as free electrons are
non-relativistic. Relativistic corrections of the SZ effect allow us
to measure the plasma temperature.

Although it has already been noticed that the use of the correct
relativistic equilibrium distribution is essential for the proper
interpretation of measurements of the SZ effect, no studies of this
problem have been performed in detail so far.\\
The relativistic kinetic theory, on which the DF derivation is
based, is still a subject of numerous debates. The relativistic
analogue of the Maxwell-Boltzman velocity DF has been proposed by
Juttner (1911). However, alternatives to a Juttner DF have been
discussed by Horwitz et al. (1989) and, recently, by Lehmann (2006)
and Dunkel \& Hanggi (2007).

In this paper, we showed how to separate the CMB distortions (caused
by the SZ effect) that are caused by a departure from the diffusive
approximation given by Kompaneets (1957) from those that are caused
by using a relativistic correct DF instead of a Maxwell-Boltzman DF.
We propose here a method based on Fourier analysis to derive a
velocity DF of electrons by using SZ observations at four
frequencies.

We found that the SZ intensity change owing to using a relativistic
correct DF instead of a Maxwell-Boltzman DF contributes a
significant part to the total relativistic corrections of the SZ
effect and is more significant at lower temperatures $\approx 5$ keV
than that at higher temperatures $\approx 15$ keV. We conclude that
the value of the SZ intensity change owing to using a relativistic
correct DF instead of a Maxwell-Boltzman DF will be much higher in
hot galaxy clusters because the value of the relativistic SZ
corrections is proportional to $T^{5/2}_{\mathrm{e}}$.

We proposed a method to derive the DF of electrons using SZ
multi-frequency observations of massive galaxy clusters with high
plasma temperatures. Using a Fourier analysis we found that the
approximate electron DF represented by six Fourier cosine terms
accurately describes the relativistic Juttner DF. By means of SZ
intensity measurements at four frequencies we showed how to derive
the approximate DF of electrons. To find a suitable sample of four
frequencies for deriving the DF of electrons, we studied different
samples of frequencies. We found that the best sample includes high
frequencies $\nu$=375, 600, 700, 857 GHz, while the worst sample is
in the frequency range 300-400 GHz because the matrix
$M_{\mathrm{\lambda} k}$, which is determined in Sect. 3, is
ill-conditioned for this frequency range. In the case of the
frequency sample of $\nu$=375, 600, 700, 857 GHz to derive a DF of
electrons the allowable uncertainties in the SZ intensities are
about three orders of magnitude larger than those found for the
frequency sample $\nu$ = 300, 320, 340, 360 GHz. Using Monte-Carlo
simulations of SZ observations with the SZ intensity observational
uncertainty of 0.1\%, we showed that it is possible to distinguish
Juttner and Maxwell-Bolzman DFs by means of these SZ observations.
Therefore, the SZ effect provides us with a promising method to
study DFs of electrons in massive galaxy clusters that contain hot
plasmas with temperatures $\simeq$ $15$ keV.

We considered in our analysis the relaxation of a system of
electrons with Coulomb interactions and conclude that the Juttner
distribution function is an appropriate approximation to the
universal (equilibrium) electron distribution in hot merging
clusters.

We applied this method to derive the DF of electrons using
multi-frequency SZ observations in galaxy clusters where hard X-ray
tails were detected. We demonstrated the ability of SZ
multi-frequency observations to derive the electron DF for the
Bullet cluster and found that a precision of SZ intensity
measurements of $\approx$0.1\% is required. This method is
independent of those proposed by Blasi et al. (2000), Shimon \&
Rephaeli (2002), and Colafrancesco et al. (2009), and can be used to
distinguish among different interpretations of the X-ray excess.
Although our method requires a higher precision of SZ observations
compared with other methods, it permits us to directly study
electron DFs rather than studying the supplementary electron
pressure caused by the presence of quasi-thermal particles. This is
an important advantage of our method.

Using the 3D hydrodynamic numerical simulations of a hot merging
galaxy cluster, we also demonstrated that the SZ spatial intensity
maps of the simulated hot merging galaxy cluster at frequencies of
375 GHz, 600 GHz, 700 GHz, and 857 GHz are different. This is
because of the important role of SZ relativistic corrections at high
frequencies. New SZ multi-frequency measurements with a high spatial
resolution should confirm our conclusion.

The next generation of SZ effect experimental techniques (as those
outlined in Colafrancesco \& Marchegiani 2010) are needed to reach
the required high sensitivity with the purpose of studying electron
DFs by means of multi-frequency SZ observations.

\acknowledgements{We are grateful to Joseph Silk and Vladimir Dogiel
for valuable suggestions and discussions and thank the referee for
valuable suggestions.}

\end{document}